\documentclass[%
preprint,
aps, pra
]{revtex4-1}
\usepackage{amsmath,amssymb}
\usepackage{graphicx}
\usepackage{dcolumn}
\usepackage{bm}
\usepackage{textcomp}
\usepackage{lipsum}
\usepackage{physics}
\usepackage{hyperref}
\usepackage{xcolor}
\hypersetup{colorlinks=true, allcolors=blue}
\usepackage{enumitem}
\usepackage{multirow}

\begin{document}

\title{Phenomenological description of the dynamics of bipartite antiferromagnets in the limit of strong exchange}


\author{Arun Parthasarathy}
\affiliation{Electrical and Computer Engineering, New York University, Brooklyn, New York 11201, USA}

\author{Shaloo Rakheja}
\affiliation{Electrical and Computer Engineering, New York University, Brooklyn, New York 11201, USA}


\date{\today}

\renewcommand{\abstractname}{}
\begin{abstract}
\vspace{10pt}
The equation of motion of the staggered order parameter is derived in a step-by-step manner from the coupled Landau-Lifshitz-Gilbert dynamics of bipartite spin moments in the limit of strong antiferromagnetic exchange coupling.
\end{abstract}

\begin{minipage}[t]{\textwidth}
    \maketitle
\end{minipage}

\section{Modeling of the free energy density}
The simplest antiferromagnet has a collinear bipartite ordering composed of two interpenetrating square lattices A and B possessing oppositely aligned moments arranged in the form of a checkerboard~\cite{RevModPhys.90.015005}. For such ordering, the thermodynamic state of the antiferromagnet under mean-field approximation is represented by the spin moment unit vectors at each sublattice $\vb{m}_\text{A}$ and $\vb{m}_\text{B}$, whose magnetization is $M_\text{s}$~\cite{kittel2004introduction}. To describe the dynamics of the antiferromagnet in response to an applied magnetic field or a spin-transfer torque at a fixed temperature $T$, the magnetic free energy of the system first needs to be modeled. There are three significant contributions to the free energy density $\mathcal{F}$ of an antiferromagnet: the exchange energy due to alignment of neighboring spins, the magnetocrystalline anisotropy energy due to spin-orbit interaction, and the Zeeman energy due to interaction with an external field.

The exchange energy is subdivided into intra-sublattice component for the spatial variation of each of the sublattice moments, and inter-sublattice component for the relative orientation between the sublattice moments. Up to first-order approximation, the exchange energy is phenomenologically expressed as~\cite{PhysRevLett.118.137201}
\begin{equation}
\mathcal{F}_\text{e} =  \frac{A^{\upuparrows}}{2}\big[ (\nabla \vb{m}_\text{A})^2 + (\nabla \vb{m}_\text{B})^2 \big] + \frac{\mathcal{J}}{2} \vb{m}_\text{A}\cdot\vb{m}_\text{B} + A^{\uparrow\downarrow}(\nabla \vb{m}_\text{A}\cdot\nabla \vb{m}_\text{B}),
\label{eq:Fe}
\end{equation}
where $A^{\upuparrows}$ is the intra-sublattice exchange spring constant (J/m), while $\mathcal{J}$ and $A^{\uparrow\downarrow}$ are the inter-sublattice exchange energy density (J/m\textsuperscript{3}) and spring constant (J/m), respectively. 

The anisotropy energy also consists of intra- and inter-sublattice contributions, which can be further expanded in terms of the coordinates of the sublattice moments depending on crystal symmetry. For biaxial symmetry, the anisotropy is dichotomized into orthogonal components corresponding to the easy ($x$ axis) and the hard ($z$ axis) directions, which is expressed up to first order as~\cite{PhysRevLett.118.137201}
\begin{equation}
\mathcal{F}_\text{a} = \frac{K_z^{\upuparrows}}{2} \big(m_{\text{A},z}^2 + m_{\text{B},z}^2 \big) - \frac{K_x^{\upuparrows}}{2} \big(m_{\text{A},x}^2 + m_{\text{B},x}^2 \big) + K_z^{\uparrow\downarrow}m_{\text{A},z}m_{\text{B},z} - K_x^{\uparrow\downarrow}m_{\text{A},x}m_{\text{B},x},
\label{eq:Fa}
\end{equation}
where $K_i^\upuparrows$ and $K_i^{\uparrow\downarrow}$ are the intra- and inter-sublattice anisotropy energy densities (J/m\textsuperscript{3}), respectively for the directions $i \in \{x,z\}$.

The Zeeman energy in the presence of an external magnetic field $H$ applied along the unit vector $\vb{h}$ is given as
\begin{equation}
\mathcal{F}_\text{Z} = -\mathcal{Z}(\vb{m}_\text{A} + \vb{m}_\text{B})\cdot\vb{h},\quad \mathcal{Z} = \mu_0 M_\text{s} H.
\label{eq:FZ}
\end{equation}

For reasons which will be clearer in the next section, let us deploy change of variables from $\vb{m}_\text{A}, \vb{m}_\text{B}$ to $\vb{m},\vb{n}$, where $\vb{m} = (\vb{m}_\text{A} + \vb{m}_\text{B})/2$ is the average net moment and $\vb{n}=(\vb{m}_\text{A} - \vb{m}_\text{B})/2$ is the staggered order parameter. Consider the case where the inter-sublattice exchange is much larger than the anisotropy, Zeeman energy and thermal effects, that is $\mathcal{J} \gg K_i^\upuparrows, K_i^{\uparrow\downarrow}, \mathcal{Z}, k_\text{B}T/V$, where $V$ is the volume of the sample. This essentially means that the characteristic length scale over which the antiferromagnetic order varies is much larger than the interatomic distance~\cite{KOSEVICH1990117}.  Under such assumption, the sublattice moments would remain almost antiparallel to one another at all times (because this configuration would have the lowest total energy), so that $\vb{m}^2 \ll \vb{n}^2 = 1 - \vb{m}^2  \lessapprox 1$.

The following relations are derived to rewrite the total free energy \eqref{eq:Fe}+\eqref{eq:Fa}+\eqref{eq:FZ} in terms of the new variables $\vb{m},\vb{n}$.
\begin{align}
(\nabla \vb{m}_\text{A})^2 + (\nabla \vb{m}_\text{B})^2 &=  \frac{1}{2}\big[ \{\nabla(\vb{m}_\text{A} + \vb{m}_\text{B})\}^2 + \{\nabla(\vb{m}_\text{A} - \vb{m}_\text{B})\}^2 \big] \nonumber \\ &= 2\big[(\nabla \vb{m})^2 + (\nabla \vb{n})^2 \big] \simeq 2(\nabla \vb{n})^2.\\
\nabla \vb{m}_\text{A}\cdot\nabla \vb{m}_\text{B} &=  \frac{1}{4}\big[ \{\nabla(\vb{m}_\text{A} + \vb{m}_\text{B})\}^2 - \{\nabla(\vb{m}_\text{A} - \vb{m}_\text{B})\}^2 \big] \nonumber \\ &= (\nabla \vb{m})^2 - (\nabla \vb{n})^2 \simeq -(\nabla \vb{n})^2.\\
\vb{m}_\text{A}\cdot\vb{m}_\text{B} &= \frac{1}{2}\big[ (\vb{m}_\text{A} + \vb{m}_\text{B})^2 - \vb{m}_\text{A}^2 - \vb{m}_\text{B}^2 \big] = 2\vb{m}^2 - 1.
\end{align}
\begin{align}
m_{\text{A},i}^2 + m_{\text{B},i}^2 &= \frac{1}{2}\big[ (m_{\text{A},i} + m_{\text{B},i})^2 + (m_{\text{A},i} - m_{\text{B},i})^2 \big] = 2\big[ m_i^2 + n_i^2\big] \simeq  2n_i^2. \\
m_{\text{A},i}m_{\text{B},i} &= \frac{1}{4}\big[ (m_{\text{A},i} + m_{\text{B},i})^2 - (m_{\text{A},i} - m_{\text{B},i})^2 \big] = m_i^2 - n_i^2 \simeq -n_i^2.
\end{align}
On substitution, the total free energy density is simplified to
\begin{equation}
\mathcal{F}(\vb{m},\vb{n}) = \mathcal{A}(\nabla\vb{n})^2 + \mathcal{J}\vb{m}^2 + \mathcal{K}_z n_z^2 -\mathcal{K}_x n_x^2 - 2\mathcal{Z}\vb{m}\cdot\vb{h},
\label{eq:Fmn}
\end{equation}
where  $\mathcal{A} = A^{\upuparrows} - A^{\uparrow\downarrow}$ and $\mathcal{K}_i =  K_i^{\upuparrows} - K_i^{\uparrow\downarrow}$ are the net parameters, all of which are greater than zero because mean-field theory requires the intra-sublattice contributions to be much larger than the inter-sublattice.

\section{Landau-Lifshitz-Gilbert equation of motion}
The coupled dynamics of the sublattice moments of the antiferromagnet is described by the phenomenological Landau-Lifshitz-Gilbert equations
\begin{align}
\dot{\vb{m}}_\text{A} &=  \frac{\gamma}{M_\text{s}}\vb{m}_\text{A}\times\fdv{\mathcal{F}}{\vb{m}_\text{A}} + \alpha\vb{m}_\text{A}\times\dot{\vb{m}}_\text{A},\label{eq:mAdot}\\
\dot{\vb{m}}_\text{B} &=  \frac{\gamma}{M_\text{s}}\vb{m}_\text{B}\times\fdv{\mathcal{F}}{\vb{m}_\text{B}} + \alpha\vb{m}_\text{B}\times\dot{\vb{m}}_\text{B}\label{eq:mBdot},
\end{align}
where overdot denotes partial derivative with respect to time $\partial_t$, $\gamma$ is the electron gyromagnetic ratio and $\alpha$ is the Gilbert damping constant. Now, the dynamics of the average net moment $\vb{m}$ and the staggered order parameter $\vb{n}$ can be written as     
\begin{align}
\dot{\vb{m}} &= \frac{1}{2}(\dot{\vb{m}}_\text{A} + \dot{\vb{m}}_\text{B}) \nonumber\\
&= \frac{\gamma}{2M_\text{s}}\left(\vb{m}_\text{A}\times\fdv{\mathcal{F}}{\vb{m}_\text{A}} + \vb{m}_\text{B}\times\fdv{\mathcal{F}}{\vb{m}_\text{B}} \right) + \frac{\alpha}{2}\left( \vb{m}_\text{A}\times\dot{\vb{m}}_\text{A} + \vb{m}_\text{B}\times\dot{\vb{m}}_\text{B} \right)\label{eq:mdot1}\\
\dot{\vb{n}} &= \frac{1}{2}(\dot{\vb{m}}_\text{A} - \dot{\vb{m}}_\text{B}) \nonumber\\
&= \frac{\gamma}{2M_\text{s}}\left(\vb{m}_\text{A}\times\fdv{\mathcal{F}}{\vb{m}_\text{A}} - \vb{m}_\text{B}\times\fdv{\mathcal{F}}{\vb{m}_\text{B}} \right) + \frac{\alpha}{2}\left( \vb{m}_\text{A}\times\dot{\vb{m}}_\text{A} - \vb{m}_\text{B}\times\dot{\vb{m}}_\text{B} \right)\label{eq:ndot1}
\end{align}

Let us rewrite the dynamics of $\vb{m},\vb{n}$ as functions of  $\vb{m},\vb{n}$ alone.
From the gradient theorem, the change in free energy 
\begin{equation}
\dd \mathcal{F} = \fdv{\mathcal{F}}{\vb{m}} \dd{\vb{m}} + \fdv{\mathcal{F}}{\vb{n}} \dd{\vb{n}},
\end{equation}
so that its functional derivatives with respect to each sublattice moment become
\begin{gather}
\fdv{\mathcal{F}}{\vb{m}_\text{A}} = \left(\fdv{\mathcal{F}}{\vb{m}}\right) \left(\pdv{\vb{m}}{\vb{m}_\text{A}}\right) + \left(\fdv{\mathcal{F}}{\vb{n}}\right) \left(\pdv{\vb{n}}{\vb{m}_\text{A}}\right) = \frac{1}{2}\left( \fdv{\mathcal{F}}{\vb{m}} + \fdv{\mathcal{F}}{\vb{n}} \right),\\
\fdv{\mathcal{F}}{\vb{m}_\text{B}} = \left(\fdv{\mathcal{F}}{\vb{m}}\right) \left(\pdv{\vb{m}}{\vb{m}_\text{B}}\right) + \left(\fdv{\mathcal{F}}{\vb{n}}\right) \left(\pdv{\vb{n}}{\vb{m}_\text{B}}\right) = \frac{1}{2}\left( \fdv{\mathcal{F}}{\vb{m}} - \fdv{\mathcal{F}}{\vb{n}} \right).
\end{gather}
The following relations are derived to replace the right-hand-side terms in Eq.~\eqref{eq:mdot1} and \eqref{eq:ndot1}.
\begin{align}
\vb{m}_\text{A}\times\fdv{\mathcal{F}}{\vb{m}_\text{A}} + \vb{m}_\text{B}\times\fdv{\mathcal{F}}{\vb{m}_\text{B}} &= (\vb{m} + \vb{n})\times\fdv{\mathcal{F}}{\vb{m}_\text{A}} + (\vb{m} - \vb{n})\times\fdv{\mathcal{F}}{\vb{m}_\text{B}} \nonumber \\
&= \vb{m}\times\left( \fdv{\mathcal{F}}{\vb{m}_\text{A}} + \fdv{\mathcal{F}}{\vb{m}_\text{B}} \right) +  \vb{n}\times\left( \fdv{\mathcal{F}}{\vb{m}_\text{A}} - \fdv{\mathcal{F}}{\vb{m}_\text{B}} \right) \nonumber \\
&= \vb{m}\times\fdv{\mathcal{F}}{\vb{m}} + \vb{n}\times\fdv{\mathcal{F}}{\vb{n}}.
\end{align}
\begin{align}
\vb{m}_\text{A}\times\fdv{\mathcal{F}}{\vb{m}_\text{A}} - \vb{m}_\text{B}\times\fdv{\mathcal{F}}{\vb{m}_\text{B}} &= (\vb{m} + \vb{n})\times\fdv{\mathcal{F}}{\vb{m}_\text{A}} - (\vb{m} - \vb{n})\times\fdv{\mathcal{F}}{\vb{m}_\text{B}} \nonumber \\
&= \vb{m}\times\left( \fdv{\mathcal{F}}{\vb{m}_\text{A}} - \fdv{\mathcal{F}}{\vb{m}_\text{B}} \right) +  \vb{n}\times\left( \fdv{\mathcal{F}}{\vb{m}_\text{A}} + \fdv{\mathcal{F}}{\vb{m}_\text{B}} \right) \nonumber \\
&= \vb{m}\times\fdv{\mathcal{F}}{\vb{n}} + \vb{n}\times\fdv{\mathcal{F}}{\vb{m}}.
\end{align}
\begin{align}
\vb{m}_\text{A}\times\dot{\vb{m}}_\text{A} + \vb{m}_\text{B}\times\dot{\vb{m}}_\text{B} &= (\vb{m}+\vb{n})\times(\dot{\vb{m}}+\dot{\vb{n}}) + (\vb{m}-\vb{n})\times(\dot{\vb{m}}-\dot{\vb{n}}) \nonumber \\
&= 2(\vb{m}\times\dot{\vb{m}} + \vb{n}\times\dot{\vb{n}}) \simeq 2\vb{n}\times\dot{\vb{n}}.
\end{align}
\begin{align}
\vb{m}_\text{A}\times\dot{\vb{m}}_\text{A} - \vb{m}_\text{B}\times\dot{\vb{m}}_\text{B} &= (\vb{m}+\vb{n})\times(\dot{\vb{m}}+\dot{\vb{n}}) - (\vb{m}-\vb{n})\times(\dot{\vb{m}}-\dot{\vb{n}}) \nonumber \\
&= 2(\vb{m}\times\dot{\vb{n}} + \vb{n}\times\dot{\vb{m}}) = 2\partial_t(\vb{m}\times\vb{n}).
\end{align}
On substitution, the coupled dynamics of $\vb{m}$ and $\vb{n}$ are obtained as
\begin{align}
\dot{\vb{m}} &= \frac{\gamma}{2M_\text{s}}\left(\vb{m}\times\fdv{\mathcal{F}}{\vb{m}} + \vb{n}\times\fdv{\mathcal{F}}{\vb{n}}\right) + \alpha\vb{n}\times\dot{\vb{n}}, \label{eq:mdot2}\\
\dot{\vb{n}} &= \frac{\gamma}{2M_\text{s}}\left(\vb{m}\times\fdv{\mathcal{F}}{\vb{n}} + \vb{n}\times\fdv{\mathcal{F}}{\vb{m}} \right) + \alpha\partial_t(\vb{m}\times\vb{n})  \label{eq:ndot2}.
\end{align}

The functional derivatives of the free energy density \eqref{eq:Fmn} are evaluated as 
\begin{align}
\fdv{\mathcal{F}}{\vb{m}} &=  \fdv{\vb{m}}(\mathcal{J}\vb{m}^2 - 2\mathcal{Z}\vb{m}\cdot\vb{h}) = 2\mathcal{J}\vb{m} - 2\mathcal{Z}\vb{h} \label{eq:partialFm}, \\[2pt]
\fdv{\mathcal{F}}{\vb{n}} &= \fdv{\vb{n}}(\mathcal{A}(\nabla\vb{n})^2 + \mathcal{K}_z n_z^2 -\mathcal{K}_x n_x^2 ) = 2\left(-\mathcal{A} \nabla^2\vb{n} + \mathcal{K}_z n_z \hat{\vb{z}} - \mathcal{K}_x n_x \hat{\vb{x}}\right), \label{eq:partialFn}
\end{align}
where the functional derivative of $(\nabla \vb{n})^2$ is evaluated as
\begin{equation}
\fdv{\vb{n}}(\nabla \vb{n})^2 = \left(\pdv{\vb{n}} - \nabla\cdot\pdv{\nabla\vb{n}}\right)(\nabla \vb{n})^2 = 0 - \nabla\cdot(2\nabla\vb{n}) = -2\nabla^2\vb{n}.
\end{equation}
On replacing the functional derivatives in Eq.~\eqref{eq:ndot2}, we get
\begin{align}
\dot{\vb{n}} &= \frac{\gamma}{M_\text{s}}\left[\vb{m}\times\left(-\mathcal{A} \nabla^2\vb{n} - \mathcal{J}\vb{n} + \mathcal{K}_z n_z \hat{\vb{z}} - \mathcal{K}_x n_x \hat{\vb{x}} \right) - \mathcal{Z}\vb{n}\times\vb{h}\right] + \alpha\partial_t(\vb{m}\times\vb{n}) 
\end{align}
Notice that the term $\mathcal{A}\nabla^2 \sim \mathcal{A}/\lambda^2 = \mathcal{K}_x$, where $\lambda$ is the characteristic width of a domain wall~\cite{krishnan2016fundamentals}. Therefore, the anisotropy terms $\mathcal{K}_i$ as well as $\mathcal{A}\nabla^2$ can be ignored compared to $\mathcal{J}$ under the premise of strong exchange. So, we arrive at the intermediate step 
\begin{equation}
\dot{\vb{n}} \approx \left(\frac{\gamma\mathcal{J}}{M_\text{s}} - \alpha \partial_t\right)(\vb{n}\times\vb{m}) - \frac{\gamma\mathcal{Z}}{M_\text{s}}\vb{n}\times\vb{h}.
\end{equation}
To further simply, notice that the temporal variations $\partial_t$ occur in a scale similar to the antiferromagnetic resonance frequency $\omega_0 = \gamma\sqrt{\mathcal{J}\mathcal{K}_x/(2M_\text{s}^2)}$~\cite{kittel2004introduction}. We find that $\alpha\partial_t \sim \alpha\omega_0 \ll \gamma\mathcal{J}/M_\text{s}$, and hence $\alpha\partial_t$ can be omitted to give the next step
\begin{equation}
\dot{\vb{n}} \approx \left(\frac{\gamma\mathcal{J}}{M_\text{s}}\right)(\vb{n}\times\vb{m}) - \frac{\gamma\mathcal{Z}}{M_\text{s}}\vb{n}\times\vb{h}. 
\end{equation}
Transposing terms and crossing with $\vb{n}$ yields the following steps. 
\begin{gather}
\mathcal{J}(\vb{n}\times\vb{m})\times\vb{n} = \left[(M_\text{s}/\gamma)\dot{\vb{n}} + \mathcal{Z}\vb{n}\times\vb{h}\right]\times\vb{n},\\
\mathcal{J}\left[-(\vb{n}\cdot\vb{m})\vb{n} + (\vb{n}\cdot\vb{n})\vb{m}\right] \simeq (M_\text{s}/\gamma)\dot{\vb{n}}\times\vb{n} + \mathcal{Z}(\vb{n}\times\vb{h})\times\vb{n},\\
\mathcal{J}\vb{m} = (M_\text{s}/\gamma)\dot{\vb{n}}\times\vb{n} + \mathcal{Z}(\vb{n}\times\vb{h})\times\vb{n},
\end{gather}
which finally result in the solution 
\begin{equation}
\vb{m} = \left[\frac{M_\text{s}}{\gamma\mathcal{J}}\dot{\vb{n}} + \frac{\mathcal{Z}}{\mathcal{J}}(\vb{n}\times\vb{h})\right]\times\vb{n} = \frac{M_\text{s}}{\gamma\mathcal{J}}(\dot{\vb{n}}\times\vb{n}) + \frac{\mathcal{Z}}{\mathcal{J}}[\vb{h}-(\vb{n}\cdot\vb{h})\vb{n}],
\label{eq:m}
\end{equation}
which indicates that the dynamics of $\vb{n}$ dictates that of $\vb{m}$, making $\vb{m}$ a slave variable. Thus in the limit of strong exchange, the problem of coupled dynamics of the sublattice moments,~\eqref{eq:mAdot} and \eqref{eq:mBdot}, is effectively reduced to solving just one equation for the staggered order parameter~\eqref{eq:mdot2}, which is why the change of variables was performed in the first place.

Now, we are left with the task of replacing $\vb{m}$ in Eq.~\eqref{eq:mdot2} with~\eqref{eq:m}, so that the final equation is only in terms $\vb{n}$ and its derivatives. We begin with the left-hand side
\begin{align}
\dot{\vb{m}} &= \frac{M_\text{s}}{\gamma\mathcal{J}}\partial_t(\dot{\vb{n}}\times\vb{n}) - \frac{\mathcal{Z}}{\mathcal{J}}\partial_t[(\vb{n}\cdot\vb{h})\vb{n}] \nonumber \\
&= \frac{M_\text{s}}{\gamma\mathcal{J}} (\ddot{\vb{n}}\times\vb{n}) - \frac{\mathcal{Z}}{\mathcal{J}}[(\dot{\vb{n}}\cdot\vb{h})\vb{n} + (\vb{n}\cdot\vb{h})\dot{\vb{n}}].
\end{align}
Using Eq.~\eqref{eq:partialFm} and \eqref{eq:partialFn}, the right-hand-side terms of Eq.~\eqref{eq:mdot2} follow
\begin{align}
&\frac{\gamma}{2M_\text{s}}\left(\vb{m}\times\fdv{\mathcal{F}}{\vb{m}} + \vb{n}\times\fdv{\mathcal{F}}{\vb{n}}\right) \nonumber \\
&= -\frac{\gamma\mathcal{Z}}{M_\text{s}}(\vb{m}\times\vb{h}) + \frac{\gamma}{M_\text{s}} \vb{n}\times(-\mathcal{A} \nabla^2\vb{n} + \mathcal{K}_z n_z \hat{\vb{z}} - \mathcal{K}_x n_x \hat{\vb{x}}) \nonumber \\
&= -\frac{\mathcal{Z}}{\mathcal{J}}(\dot{\vb{n}}\times\vb{n})\times\vb{h} + \frac{\gamma\mathcal{Z}^2}{M_\text{s}\mathcal{J}}(\vb{n}\cdot\vb{h})(\vb{n}\times\vb{h}) + \frac{\gamma}{M_\text{s}} \vb{n}\times(-\mathcal{A} \nabla^2\vb{n} + \mathcal{K}_z n_z \hat{\vb{z}} - \mathcal{K}_x n_x \hat{\vb{x}}) \nonumber \\
&= \frac{\mathcal{Z}}{\mathcal{J}}\left[ (\vb{n}\cdot\vb{h})\dot{\vb{n}} - (\dot{\vb{n}}\cdot\vb{h})\vb{n} \right] +  \frac{\gamma\mathcal{Z}^2}{M_\text{s}\mathcal{J}}(\vb{n}\cdot\vb{h})(\vb{n}\times\vb{h}) + \frac{\gamma}{M_\text{s}} \vb{n}\times(-\mathcal{A} \nabla^2\vb{n} + \mathcal{K}_z n_z \hat{\vb{z}} - \mathcal{K}_x n_x \hat{\vb{x}})
\end{align}
On substituting the left- and right-hand-side terms, we arrive at
\begin{align}
\frac{M_\text{s}}{\gamma\mathcal{J}} (\vb{n}\times\ddot{\vb{n}}) +  &\frac{2\mathcal{Z}}{\mathcal{J}}(\vb{n}\cdot\vb{h})\dot{\vb{n}} +  \frac{\gamma\mathcal{Z}^2}{M_\text{s}\mathcal{J}}(\vb{n}\cdot\vb{h})(\vb{n}\times\vb{h}) + \nonumber \\
&\frac{\gamma}{M_\text{s}} \vb{n}\times(-\mathcal{A} \nabla^2\vb{n} + \mathcal{K}_z n_z \hat{\vb{z}} - \mathcal{K}_x n_x \hat{\vb{x}}) + \alpha\vb{n}\times\dot{\vb{n}} = 0,
\end{align}
which is condensed as
\begin{align}
\vb{n}\times\bigg[\ddot{\vb{n}} + \frac{2\gamma\mathcal{Z}}{M_\text{s}}(\dot{\vb{n}}\times\vb{h}) + &\frac{\gamma\mathcal{J}}{M_\text{s}}\alpha\dot{\vb{n}} + \frac{\gamma^2\mathcal{Z}^2}{M_\text{s}^2}(\vb{n}\cdot\vb{h})\vb{h} \nonumber\\ 
& \frac{\gamma^2\mathcal{J}}{M_\text{s}^2}(-\mathcal{A} \nabla^2\vb{n} + \mathcal{K}_z n_z \hat{\vb{z}} - \mathcal{K}_x n_x \hat{\vb{x}})   \bigg] = 0.
\label{eq:eomnd}
\end{align}

It is easy to check that with the following substitutions
\begin{gather}
t= \tau/(\gamma\mu_0 M_\text{s}),\quad \mathcal{A}\nabla^2=\mathcal{K}_x(\lambda\nabla)^2=\mathcal{K}_x\nabla^2_\lambda 
\nonumber \\
\mathcal{J} = (\mu_0 M_\text{s}^2)h_{\mathcal{J}},\quad \mathcal{K}_i = (\mu_0 M_\text{s}^2)h_{\mathcal{K}_i}, \quad \mathcal{Z} = (\mu_0 M_\text{s}^2)h_{\mathcal{Z}},
\end{gather}
equation~\eqref{eq:eomnd} assumes the nondimensionalized form
\begin{align}
\vb{n}\times\bigg[\frac{\ddot{\vb{n}}}{h_\mathcal{J}} + &\frac{2h_\mathcal{Z}}{h_\mathcal{J}}(\dot{\vb{n}}\times\vb{h}) + \alpha\dot{\vb{n}} - h_{\mathcal{K}_x}\nabla_\lambda^2\vb{n} + \nonumber\\
&h_{\mathcal{K}_z} n_z \hat{\vb{z}} - h_{\mathcal{K}_x} n_x \hat{\vb{x}} + \frac{h_\mathcal{Z}^2}{h_\mathcal{J}}(\vb{n}\cdot\vb{h})\vb{h} \bigg] = 0.
\end{align}
There are two interesting takeaways from the equation of motion: (a) the order parameter exhibits an inertial timescale $\tau\sim 1/(\alpha h_\mathcal{J})$ because of the second order time derivative; (b) antiferromagnets are resilient to stray magnetic fields, since the Zeeman terms are diminished by the strong exchange.

Let us include the local spin-transfer torque~\cite{Gomonay2010}
\begin{equation}
\partial_t\vb{m}_\text{A,B}\big|_\text{STT} = \frac{\gamma J_\mu}{M_\text{s}}\vb{m}_\text{A,B}\times(\vb{m}_\text{A,B}\times\boldsymbol{\mu}),
\end{equation}
to the right-hand side of Eq.~\eqref{eq:mAdot} and \eqref{eq:mBdot}, where $J_\mu$ is the local spin-polarized current density (J/m\textsuperscript{3}) and $\boldsymbol{\mu}$ is the unit vector along the direction of spin polarization. It can be shown that in the final equation of motion of the order parameter
\begin{align}
\vb{n}\times\bigg[\frac{\ddot{\vb{n}}}{h_\mathcal{J}} + &\frac{2h_\mathcal{Z}}{h_\mathcal{J}}(\dot{\vb{n}}\times\vb{h}) + \alpha\dot{\vb{n}} - h_{\mathcal{K}_x}\nabla_\lambda^2\vb{n} + \nonumber\\
&h_{\mathcal{K}_z} n_z \hat{\vb{z}} - h_{\mathcal{K}_x} n_x \hat{\vb{x}} + \frac{h_\mathcal{Z}^2}{h_\mathcal{J}}(\vb{n}\cdot\vb{h})\vb{h} + j_\mu(\vb{n}\times\boldsymbol{\mu}) \bigg] = 0,
\label{eq:eomsttnd}
\end{align}
the spin-transfer torque appears as an add-on term inside the square brackets (similar to how the Gilbert damping term shows up), where $j_\mu = J_\mu/(\mu_0 M_\text{s}^2)$.

\section{Conclusion}
The equation of motion~\eqref{eq:eomsttnd} we have derived for the staggered order parameter is consistent with that of Ref.~\onlinecite{KOSEVICH1990117}, while including both Gilbert damping and spin-transfer torque phenomenon. It is also consistent with the equation of motion of Ref.~\onlinecite{PhysRevLett.118.137201}, which uses the Lagrangian approach (note there is an extra factor of 1/4 in the Berry phase part of the Lagrangian).

\begin{acknowledgments}
This work was supported in part by the Semiconductor Research Corporation (SRC) and the National Science Foundation (NSF) through ECCS 1740136. S. Rakheja also acknowledges the funding support from the MRSEC Program of the National Science Foundation under Award Number DMR-1420073. We would like to thank Dr. A. Qaiumzadeh for useful email discussions and pointing out a vital error in our derivation.
\end{acknowledgments}

\bibliography{refs}

\end{document}